\documentclass[12pt]{article}
\usepackage{amsmath,amsfonts,latexsym,graphicx,amssymb}
\date{}
\def\la{\langle\,}
\def\r{\,\rangle}
\newcommand{\eeq}{\end{eqnarray}}
\newcommand{\beq}{\begin{eqnarray}}
\newcommand{\call}{{\cal  L}}
\newcommand{\calp}{{\cal  P}}

\newcommand{\calh}{\cal H}
\def\con{{}_{\_\rule{-1pt}{0pt}\_}
\rule{-2pt}{0pt}\raise1.5pt\hbox{$\mid$}\hspace{2pt}}

\title{Resonant states and classical damping}
\author{Dariusz Chru\'sci\'nski \\
 Institute of Physics, Nicholas Copernicus University\\
 ul. Grudzi\c{a}dzka 5/7, 87-100 Toru\'n, Poland\\
 e-mail: darch@phys.uni.torun.pl}

\begin{document}

\maketitle

\begin{abstract}
Using Koopman's approach to classical dynamical systems we show
that the classical damping may be interpreted as appearance of
resonant states  of the corresponding Koopman's operator. It turns
out that simple classical damped systems give rise to discrete
complex spectra. Therefore, the corresponding generalized
eigenvectors may be  interpreted as  classical resonant states.

\end{abstract}

\section{Introduction}

It is widely believed that resonant states in quantum mechanics
are responsible for the irreversible dynamics of physical systems
(see e.g. \cite{Bohm}). In this paper we address a similar problem
in classical mechanics. Usually resonant states are eigenstates of
the Hamiltonian corresponding to complex eigenvalues
``$E+i\Gamma$'' with $E,\Gamma \in \mathbb{R}$. $E$ and $\Gamma$
are called resonance energy and resonance width, respectively.
According to Breit-Wigner formula $1/\Gamma$ measures the lifetime
of the resonant state. Although quantum mechanics teaches us that
a self-adjoint operator has real eigenvalues this is true only for
proper eigenvectors, that is, eigenvectors belonging to the
Hilbert space. It is well known that generalized eigenvectors
(which do not belong to the Hilbert space) may correspond to
complex (generalized) eigenvalues. The proper mathematical
language to describe generalized eigenvectors is the rigged
Hilbert space (or the Gelfand triplet) \cite{RHS1},\cite{RHS2}:
\begin{equation}\label{triplet}
  D \ \subset \ {\calh} \ \subset \ D^* \ ,
\end{equation}
where $D$ is a dense nuclear subspace in $\calh$ and $D^*$ is a
dual of $D$. Using convenient Dirac notation $  \la \phi|\psi\r $
to denote the action of $\psi \in D^*$ on a  test function
$\phi\in D$, we call an  element $\psi$  a generalized eigenvector
of a self-adjoint operator $A$ in $\calh$ if
\begin{equation}\label{}
  \la A\phi|\psi\r = \lambda \la\phi|\psi\r \ ,
\end{equation}
for any test function $\phi \in D$.

Recently, it was observed by Kossakowski \cite{Kossak1},
\cite{Kossak} that quantization of classical damped systems leads
immediately to resonant states.  Let us recall that usually
generalized eigenvectors correspond to continuous spectrum, e.g.
plane waves for a free particle. Interestingly, studying
 quantized damped
harmonic oscillator and the quantized toy model of the classical
damped system described by
\begin{equation}\label{system}
  \dot{x} = - \gamma x \ ,
\end{equation}
where $\gamma > 0$ denotes a damping constant, Kossakowski found
that both spectra are discrete and complex. In the present paper
we show that  the original classical damped systems may be treated
in perfect analogy to their quantum counterparts. Moreover, the
characteristic exponential decay
\begin{equation}\label{}
  x(t) \ \propto\ e^{-\gamma t}
\end{equation}
may be interpreted as appearance  of a  classical resonant state.
The methods of Hilbert spaces are not reserved to quantum
mechanics. As is well known classical Hamiltonian systems may be
investigated in a very similar way. It was already observed by
Koopman \cite{Koopman} that the Hamiltonian dynamics gives rise to
one-parameter unitary group in the Hilbert space of square
integrable function on the classical phase space (we review the
Koopman's approach in Section~\ref{sec-Koopman}). Therefore, the
spectrum of the corresponding self-adjoint generator should
contain basic physical information about the classical system. It
turns out that the corresponding Koopman's operators have very
similar spectra as their quantum counterparts. Both spectra  are
discrete and complex. Their imaginary parts are responsible for
the damping phenomena on a purely classical level.

\section{Koopman's approach}
\label{sec-Koopman}

Consider a classical Hamiltonian system defined on a symplectic
manifold $({\cal P},\Omega)$. Recall, that $\Omega$ gives rise to
the Poisson bracket in the space of functions on $\calp$:
\begin{equation}\label{}
  \{F,G\} := \Omega(\call_F,\call_G)\ ,
\end{equation}
where $\call_F$ and $\call_G$ are Hamiltonian vector fields
corresponding to $F$ and $G$, respectively. The classical phase
space $\cal P$ carries the canonical volume form $d\mu$ which is
the exterior product of $n$ copies of $\Omega$ (where $2n = {\rm
dim}\calp$):
\begin{equation}\label{}
  d\mu := \Omega \wedge \ldots \wedge \Omega\ .
\end{equation}
One usually calls $d\mu$ the  Liouville measure on $\calp$. In
local canonical coordinates $(x^1, \ldots , x^n, p_1, \ldots ,
p_n)$ on $\cal P$ one has $d\mu = \prod_{i=1}^n dx^i dp_i$).
Define the  Hilbert $L^2({\cal P},d\mu)$ space of square
integrable function on $\cal P$:
\[   L^2({\cal P},d\mu) = \Big\{\, f : {\cal P} \longrightarrow
\mathbb{C}\ \Big|\ \int |f|^2d\mu < \infty\, \Big\}  \ ,  \]
 together with a Hermitian product:
\[   \la f | g \r = \int \overline{f}g d\mu   \ , \]
for any $f,g \in L^2({\cal P},d\mu)$.
 It was observed by Koopman
\cite{Koopman} (see also \cite{Arnold} and \cite{Reed} for the
review of Koopman's approach) that a canonical transformation on
$\cal P$, i.e. a map
\[   \varphi\ :\ \cal P\ \longrightarrow\ \cal P\ ,  \]
leaving a Poisson bracket invariant, gives rise to a unitary
operator on $L^2({\cal P},d\mu)$:
\begin{equation}
   U(f) := \varphi^* f = f \circ \varphi \ ,
\end{equation}
for any $f \in L^2({\cal P},d\mu)$. Indeed, $U$ is linear and maps
$L^2({\cal P},d\mu)$ into itself. Moreover, $U$ is bijection, i.e.
 any function $g\in L^2({\cal P},d\mu)$ may be written as $g =
 U(f)$ for some $f \in L^2({\cal P},d\mu)$ --- clearly, $f = g
 \circ \varphi^{-1}$. Finally, $U$ defines an isometry:
\begin{equation}\label{}
  ||U(f)||^2 =  \int_{{\cal P}} |f(\varphi(x))|^2 d\mu(x) =
\int_{{\cal P}} |f(x)|^2 d\mu(x) = ||f||\ ,
\end{equation}
due to the $\varphi$--invariance of $d\mu$.

Now, let $H \in C^\infty(\calp)$ denotes the  Hamiltonian and let
$\Phi_t$ stand for the corresponding Hamiltonian flow on $\cal P$,
i.e.
\begin{equation}\label{}
  (x(t),p(t)) = \Phi_t(x_0,p_0)  \ , \ \ \ \ {\rm for} \ t\in
  (a,b)\ .
\end{equation}
Clearly, for any $t\in (a,b)$,  $\Phi_t$ defines a canonical
transformation on $\cal P$. Recall, that if $F$ is a function on
$\cal P$, then the corresponding Hamiltonian vector field, denoted
by $ \call_{F}$, is defined by:
\[   \call_{F}(G) := \{ G, F\}\ ,  \]
for any function $G$ on $\cal P$. It turns out \cite{Povzner} that
if the Hamiltonian flow $\Phi_t$ is complete, that is, it is
defined for all $t \in \mathbb{R}$, then
\begin{equation}\label{}
  U(t)(f) = f \circ \Phi_t\ ,
\end{equation}
defines a continuous one-parameter group of unitary transformation
on $L^2({\cal P},d\mu)$. Moreover, one shows that
\begin{equation}\label{}
  U(t) = e^{ \call_{H}t} \ ,
\end{equation}
and the Stone's theorem implies that $i \call_{H}$ is
self-adjoint.

\section{Examples of classical spectra}

To get a feeling how the classical spectra of $i\call_H$ may look
like let us consider two simple examples. As a first example
consider a free particle in $ \mathbb{R}^3$. The corresponding
phase space $\calp= \mathbb{R}^6$ and the Hamiltonian is given by
$H({\bf x},{\bf p})={\bf p}^2/2m$. Therefore, the Koopman's
operator $i\call_H$ reads:
\begin{equation}\label{}
  i\call_H = \frac{i}{2m}\, {\bf p} \cdot \nabla \ ,
\end{equation}
and it is evidently self-adjoint. The corresponding eigenvalue
problem is easy to solve. One finds
\begin{equation}\label{}
  i\call_H \, f_{\bf k}({\bf x},{\bf p}) = \frac{k^2}{2m} \,  f_{\bf k}({\bf x},{\bf
  p})\ ,
\end{equation}
where the eigenvector $ f_{\bf k}({\bf x},{\bf p})$ is given by:
\begin{equation}\label{}
   f_{\bf k}({\bf x},{\bf p}) = \delta({\bf p}- {\bf k})\,
   e^{-i{\bf k}{\bf x}} \ .
\end{equation}
Hence, the spectrum is continuous, and the eigenvectors $f_{\bf
k}$ do not belong to the Hilbert space $L^2( \mathbb{R}^6)$.
However, we may consider the following Gelfand triplet
\begin{equation}\label{}
  {\cal S}( \mathbb{R}^6) \ \subset \ L^2( \mathbb{R}^6) \ \subset \ {\cal S}^*(
  \mathbb{R}^6)\ ,
\end{equation}
where by ${\cal S}( \mathbb{R}^n)$ the Schwartz space of rapidly
decreasing  functions on $ \mathbb{R}^n$. The elements of the dual
space ${\cal S}^*( \mathbb{R}^n)$  are called tempered
distributions (see e.g. \cite{Schwartz},\cite{Yosida}). Clearly,
the generalized eigenvectors $f_{\bf k}$ are tempered
distributions, i.e. elements from the dual space ${\cal S}^*(
\mathbb{R}^6)$. Note, that
\begin{equation}\label{}
  \psi_{\bf k}({\bf x}) = \int d{\bf p}\, f_{\bf k}({\bf x},{\bf
  p}) = e^{-i{\bf k}{\bf x}} \ ,
\end{equation}
defines a generalized eigenvector, i.e. element from ${\cal S}^*(
\mathbb{R}^3)$,  for the corresponding quantum system with
$\widehat{H} = -(\hbar^2/2m) \triangle$, that is,
\begin{equation}\label{}
  \widehat{H} \, \psi_{\bf k} = \frac{\hbar^2 k^2}{2m}\, \psi_{\bf k} \ .
\end{equation}
Therefore, there is the direct correspondence between the spectrum
of $\widehat{H}$ and that of $i\call_H$.

Our next example is a harmonic oscillator. In suitable coordinates
the oscillator Hamiltonian is given by:
\begin{equation}\label{}
  H_{\rm osc} = \frac{\omega}{2} (x^2 + p^2) \ ,
\end{equation}
and therefore, the corresponding Koopman's operator reads:
\begin{equation}\label{}
  i\call_{H_{\rm osc}} = i\omega \left(p\partial_x - x\partial_p\right)\ .
\end{equation}
Clearly, the Hamiltonian flow $\Phi^{\rm osc}_t$ is complete, and
hence $i\call_{H_{\rm osc}}$  defines a self-adjoint operator on
$L^2( \mathbb{R}^2)$. Introducing complex coordinates
$(z=x+ip,\overline{z}=x-ip)$ one obtains:
\begin{equation}\label{}
  H_{\rm osc} = \frac{\omega}{2}\, |z|^2 \ ,
\end{equation}
and
\begin{equation}\label{Koopman-H-osc}
i\call_{H_{\rm osc}} = \omega \left( z\partial - \overline{z}
\overline{\partial}\right) \ ,
\end{equation}
where we use the standard notation: $\partial = \partial/\partial
z$ and $\overline{\partial} = \partial/\partial \overline{z}$.
One immediately finds the corresponding eigenvectors:
\begin{equation}\label{}
\left( z\partial - \overline{z} \overline{\partial}\right)\, z^n
\overline{z}^m = (n-m) \, z^n \overline{z}^m \ ,
\end{equation}
where $n,m$ are non-negative integers. Clearly the functions $z^n
\overline{z}^m$ do not belong to $L^2( \mathbb{R}^2)$, however
this deficiency may be easily cured. Observe, that any function of
$|z|$ belongs to the kernel of $i\call_{H_{\rm osc}}$. Define the
normalized functions:
\begin{equation}\label{f-nm}
  f_{nm}(z,\overline{z}) =  \frac{ z^n \overline{z}^m}{\sqrt{\pi(n+m)!}} \ e^{-
  |z|^2/2} \ .
\end{equation}
 This normalization leads to the following formula:
\begin{equation}\label{}
   \la f_{nm}|f_{kl}\r =
   \frac{(m+k)!}{\sqrt{(n+m)!(k+l)!}}\, \delta_{n-m,k-l}\ .
\end{equation}
Obviously
\begin{equation}\label{}
  i\call_{H_{\rm osc}}\, f_{nm} = \omega(m-n)\, f_{nm} \ ,
\end{equation}
that is, the spectrum of the Koopman's operator reads:
\begin{equation}\label{Koopman-Osc}
  {\rm Spec}(i\call_{H_{\rm osc}}) = \{ \ \omega l \ | \ l \in
  \mathbb{Z}\ \}\ .
\end{equation}
Let us compare the spectrum of $i\call_{H_{\rm osc}}$ with the
spectrum of $\widehat{H}_{\rm osc}$ acting on $L^2( \mathbb{R})$.
It is convenient to use the Bargmann representation
\cite{Bargmann}, i.e. consider a Hilbert space of holomorphic
functions $\psi=\psi(z)$ equipped with the following scalar
product:
\begin{equation}\label{B-product}
  \la \psi|\phi\r_{\rm B} = \int \overline{\psi}(z) \phi(z) e^{-|z|^2}\,
  dzd\overline{z} \ .
\end{equation}
Introducing standard creation and anihilation operators:
\begin{eqnarray}\label{a-a*}
  \widehat{a} = \frac{ \widehat{x} + i\widehat{p} }{\sqrt{2\hbar}} \ ,
  \hspace{1cm}
 \widehat{a}^* = \frac{ \widehat{x} - i\widehat{p} }{\sqrt{2\hbar}} \ ,
\end{eqnarray}
one obtains
\begin{equation}\label{H-aa*}
  \widehat{H}_{\rm osc} = \hbar\omega \left( a^*a + \frac 12 \right) \
  .
\end{equation}
Moreover, the well known commutation relation
\begin{equation}\label{}
  [ \widehat{a},\widehat{a}^*] = 1\ ,
\end{equation}
has the following holomorphic representation:
\begin{equation}\label{}
  \widehat{a}\psi(z) = \partial\psi(z)\ \ \ \ \ {\rm and} \ \ \ \ \ \widehat{a}^*\psi(z) =
  z\psi(z)\ ,
\end{equation}
and therefore, due to (\ref{H-aa*}), one obtains for the
oscillator Hamiltonian:
\begin{equation}\label{H-z-dz}
  \widehat{H}_{\rm osc} =  \hbar\omega \left( z\partial + \frac 12 \right) \
  .
\end{equation}
Now the similarity between classical formula (\ref{Koopman-H-osc})
and the quantum one (\ref{H-z-dz}) is evident. Recall that
$\widehat{H}_{\rm osc}$ acts on holomorphic functions, i.e.
functions of $z$ only, whereas the classical counterpart
$i\call_{H_{\rm osc}}$ acts on functions depending both on $z$ and
$\overline{z}$. Therefore, we have an additional term $
\overline{z}\overline{\partial}$ in (\ref{Koopman-H-osc}). The
normalized eigenvectors of $\widehat{H}_{\rm osc}$ are given by:
\begin{equation}\label{}
  \psi_n(z) = \frac{z^n}{\sqrt{\pi n!}}\ , \ \ \ \ n=0,1,2,\ldots
  \ ,
\end{equation}
and hence
\begin{equation}\label{Spec-Osc}
  {\rm Spec}(\widehat{H}_{\rm osc}) = \left\{ \ \hbar \omega \left( n +
  \frac 12 \right) \ \Big| \ n=0,1,2,\ldots \ \right\} \ .
\end{equation}
Note that the weight factor $e^{-|z|^2}$ present in the definition
of the scalar product (\ref{B-product}) is already contained in
the classical eigenvectors $f_{nm}$, cf. (\ref{f-nm}). Clearly, we
have the direct correspondence:
\begin{equation}\label{}
  f_{n0} \longleftrightarrow \psi_n \ \ \ \ \ {\rm and} \ \ \ \ \
  f_{0n} \longleftrightarrow \overline{\psi}_n \ ,
\end{equation}
and hence holomorphic $f_{n0}$ are responsible for the positive
part of the spectrum, whereas anti-holomorphic $f_{0n}$ for the
negative part.

\section{A toy model for a damped motion}

Now, let us turn to classical damped systems.  Consider the
simplest one:
\begin{equation}\label{damp-1}
  \dot{x} = - \gamma x \ .
\end{equation}
Clearly this system is not Hamiltonian. However, it is well known
(cf. e.g. \cite{Pontr}) that  any dynamical system  may be
rewritten in a Hamiltonian form. Consider a dynamical system on
$n$-dimensional configuration space $Q$:
\begin{equation}\label{dotx-X}
  \dot{x} = X(x)\ ,
\end{equation}
where $X$ is a vector field on $Q$. Now, define the following
Hamiltonian on the cotangent bundle ${\cal P}=T^*Q$:
\begin{equation}\label{}
  H(\alpha_x) := \alpha_x(X(x))\ ,
\end{equation}
where $\alpha_x \in T_x^*Q$. Using canonical coordinates
$(x^1,\ldots,x^n,p_1,\ldots,p_n)$ one obtains:
\begin{equation}\label{}
  H(x,p) = \sum_{k=1}^n p_kX^k(x)\ ,
\end{equation}
where $X^k$ are components of $X$ in the coordinate basis
$\partial/\partial x^k$. The corresponding Hamilton equations take
the following form:
\begin{eqnarray}\label{HAM-1}
  \dot{x}^k &=& \{ x^k,H\} = X^k(x) \ , \\
  \dot{p}_k &=& \{ p_k,H\} = - \sum_{l=1}^n p_l \frac{\partial
  X^l(x)}{\partial x^k} \ ,
\end{eqnarray}
for $k=1,\ldots,n$. In the above formulae $\{\ , \ \}$ denotes the
canonical Poisson bracket on $T^*Q$:
\begin{equation}\label{}
  \{F,G\} = \sum_{k=1}^n \left( \frac{\partial F}{\partial x^k}
   \frac{\partial G}{\partial p_k} - \frac{\partial G}{\partial x^k}
   \frac{\partial F}{\partial p_k}  \right) \ .
\end{equation}
Clearly, the formulae (\ref{HAM-1}) reproduce our initial
dynamical system (\ref{dotx-X}) on $Q$.

Let us apply the above procedure to the damped system
(\ref{damp-1}). One obtains for the Hamiltonian
\begin{equation}\label{}
  H(x,p) = - \gamma xp \ ,
\end{equation}
and hence the corresponding Hamilton equations
\begin{eqnarray}\label{}
\dot{x} &=& - \gamma x \ , \\ \dot{p} &=& \gamma p \ ,
\end{eqnarray}
give rise to the following Hamiltonian flow on $ \mathbb{R}^2$:
\begin{equation}\label{}
  \Phi_t(x,p) = (e^{-\gamma t}x,e^{t\gamma}p)\ .
\end{equation}
Evidently, $\Phi_t$ is complete, and hence the Koopman's operator
\begin{equation}\label{}
  i\call_{H} = i\gamma \left(p\partial_p - x\partial_x\right) \ .
\end{equation}
is  self-adjoint on $L^2( \mathbb{R}^2)$.\footnote{This may be
easily checked by formal integration by parts:
\begin{eqnarray}\label{}
 \la f,i\call_H\, g\r &=& i\gamma \int \overline{f} (p\partial_p - x\partial_x)
 \, g\, d\mu \nonumber \\ &=& - i\gamma \int \left[ p(\partial_p\overline{f})g +
 fg - x(\partial_x\overline{f})g - fg \right]d\mu = \la i\call_H \,
 f,g\r \ ,
\end{eqnarray}
assuming that $f$ and $g$ vanish sufficiently fast for $|x|,|p|
\longrightarrow \infty$.} Note that
\begin{equation}\label{}
  i\call_H g_{nm} = i\gamma (m-n)\,
  g_{nm}\ ,
\end{equation}
where
\begin{equation}\label{}
g_{nm}(x,p) = x^n p^m\ ,
\end{equation}
and $n,m$ are non-negative integers. Interestingly, the
generalized eigenvectors $g_{mn} \in {\cal S}^*(\mathbb{R}^2)$
correspond to purely imaginary discrete spectrum
\begin{equation}\label{Spec-c}
  {\rm Spec}(i\call_H) = \{\ i\gamma\, k \ |\ k \in \mathbb{Z}\ \}\ .
\end{equation}
Therefore, we call $g_{nm}$ the {\em classical resonant states}.
Let us observe that
\begin{equation}\label{}
  g_{10} = x \  \ \ \ \ {\rm and}\ \ \ \ \ g_{01} =
  p\ ,
\end{equation}
therefore
\begin{equation}\label{}
  i\call_{H}x = - i\gamma x \ \ \ \ \ {\rm and} \ \ \ \ \
  i\call_{H}p = i\gamma p \ ,
\end{equation}
that is, $x$ and $p$ are generalized eigenvectors of $i\call_{H}$
corresponding to imaginary eigenvalues $-i\gamma$ and $i\gamma$,
respectively. This shows that the damping in a classical system
$\dot{x} = -\gamma x$  corresponds to imaginary (generalized)
eigenvalues of Koopman's Hamiltonian $ i\call_{H}$:
\begin{equation}\label{}
  U(t) x = e^{ \call_{H}t}x = e^{-\gamma t} x \ \ \ \ \ {\rm and}
  \ \ \ \ \
U(t) p = e^{ \call_{H}t}p = e^{\gamma t} p \ ,
\end{equation}
or, in general
\begin{equation}\label{}
  U(t)f(x,p) = f( e^{-\gamma t} x,  e^{\gamma t} p) \ .
\end{equation}
There is a close relation between spectrum of a classical
Koopman's operator $i\call_H$ and the spectrum of the
corresponding quantum Hamiltonian \cite{Kossak}:
\begin{equation}\label{}
  \widehat{H} = - \frac{\gamma}{2} \left( \widehat{x}\widehat{p} +
  \widehat{p}\widehat{x} \right) \ .
\end{equation}
Let $\phi$ and $\psi$ be  solutions of
\begin{equation}\label{}
  \widehat{x}\, \phi = 0 \ \ \ \ \ {\rm and} \ \ \ \ \
 \widehat{p}\, \psi = 0\ ,
\end{equation}
respectively.  Clearly,  up to  non-important constants
\begin{equation}\label{}
\phi(x) = \delta(x) \ \ \ \ \ {\rm and} \ \ \ \ \ \psi(x)=1\ .
\end{equation}
 Note, that
\begin{equation}\label{H-phi}
  \widehat{H}\phi = - i\hbar\frac{\gamma}{2}\,\phi\ \ \ \ \ \ {\rm and} \ \ \ \ \
\widehat{H}\psi =  i\hbar\frac{\gamma}{2}\,\psi \ .
\end{equation}
 Now, define
\begin{equation}\label{}
  \phi_n := {\widehat{p}}^n\phi  \ \ \ \ \ {\rm and} \ \ \ \ \
 \psi_n := {\widehat{x}}^n\psi\ ,
\end{equation}
with $n=1,2,\ldots,$ i.e.
\begin{equation}\label{}
  \phi_n(x) = (-i\hbar)^n \delta^{(n)}(x) \ \ \ \ \ {\rm and} \ \ \ \ \
  \psi_n(x) = x^n \ .
\end{equation}
One easily finds that
\begin{equation}\label{}
 \widehat{H}\phi_n = - i\hbar\gamma\left(n + \frac{1}{2}\right)\,\phi_n\
\ \ \ \ \ {\rm and} \ \ \ \ \ \widehat{H}\psi_n = +
i\hbar\gamma\left(n + \frac{1}{2}\right)\,\psi_n\ .
\end{equation}
This shows that $ \widehat{H}$ has a purely imaginary discrete
spectrum:
\begin{equation}\label{Spec-q}
  {\rm Spec}( \widehat{H}) = \left\{ \  i\hbar\gamma\left(k +
  \frac{1}{2}\right)\ \Big| \ k \in \mathbb{Z} \ \right\} \ .
\end{equation}
which is closely related to the {\em classical spectrum}
(\ref{Spec-c}). It should be stressed that (\ref{Spec-c}) is not a
classical limit of (\ref{Spec-q}) (actually, the classical limit
of (\ref{Spec-q}) contains only zero eigenvalue). These two
spectra corresponds to different operators defined in different
spaces: $L^2( \mathbb{R}^2)$ and $L^2( \mathbb{R})$, respectively.
Moreover, let us observe that there is direct correspondence
between the generalized eigenvectors $g_{mn}$ and $\phi_n,\psi_n$:
\begin{equation}\label{}
 x^n = g_{n0} = \psi_n(x) \ \ \ \ {\rm and} \ \ \ \ p^n =
 g_{0n} = \int e^{ipx/\hbar} \phi_n(x)\, dx \ .
\end{equation}

\par
\noindent {\bf Remark 1}. There is a striking relation between
classical oscillator spectrum (\ref{Koopman-Osc}) and the spectrum
of damped system (\ref{Spec-c}), i.e. a damped system has the same
spectrum as a classical oscillator with imaginary frequency
$\omega=i\gamma$. Let us observe that performing a canonical
transformation $(x,p) \longrightarrow (X,P)$:
\begin{equation}\label{}
  x =  \frac{1}{\sqrt{2}} (X+  P) \ \ \ \ \ {\rm
  and} \ \ \ \ \ p= \frac{1}{\sqrt{2}} (X -  P) \ ,
\end{equation}
one obtains
\begin{equation}\label{H-XP}
  H = - \gamma xp = \frac{\gamma}{2} ( P^2 -  X^2) \ ,
\end{equation}
i.e. in the new variables $(X,P)$, $H$ corresponds formally to the
harmonic oscillator with  $\omega = \pm i\gamma$. This
correspondence may be easily seen  if one applies the following
linear operator \cite{Kossak}:
\begin{equation}\label{}
  {V}_\lambda := e^{i\lambda\call_{XP}} =
  e^{i\lambda(P\partial_P - X\partial_X)} \ ,
\end{equation}
with $\lambda \in \mathbb{R}$. Clearly,
\begin{equation}\label{}
 {V}_\lambda X = e^{-i\lambda} X \ \ \ \ \ {\rm and} \ \ \ \
 \ {V}_\lambda P = e^{i\lambda} P \ ,
\end{equation}
and hence $V_\lambda$ defines a complex scaling. The above
formulae imply:
\begin{equation}\label{}
{V}_\lambda ( P^2 -  X^2) = e^{2i\lambda} P^2 - e^{-2i\lambda}X^2
= e^{2i\lambda} (P^2 -  e^{-4i\lambda} X^2 ) \ ,
\end{equation}
so, if $ \lambda = \pm \pi/4$, then
\begin{equation}\label{}
  {V}_{\pm\pi/4} \Big[ \frac{\gamma}{2} ( P^2 -  X^2)\Big] = \pm \frac{i\gamma}{2} (P^2 +
 X^2) \ ,
\end{equation}
that is, both systems are related by a complex scaling
$V_{\pm\pi/4}$. Evidently, the same relation may be established
between corresponding quantum systems and their spectra
(\ref{Spec-Osc}) and (\ref{Spec-q}).

\par
\noindent {\bf Remark 2.} Let us note that if we replace
(\ref{system}) by a more general equation
\begin{equation}\label{system-n}
  \dot{x} = -\gamma x^n \ ,
\end{equation}
with $n > 1$, then the corresponding Hamiltonian system
reproducing (\ref{system-n}) is given by
\begin{equation}\label{}
  H= - \gamma x^np\ .
\end{equation}
The solution of (\ref{system-n}) reads:
\begin{equation}\label{}
  x(t) = \frac{1}{[(n-1)\gamma t ]^{1/n-1}} + c \ ,
\end{equation}
with $c = {\rm const.}$, and evidently it is not complete.
Therefore,
\begin{equation}\label{}
  U(t) = e^{t\call_H}
\end{equation}
is not defined for all $t\in \mathbb{R}$, and hence,
 Hamiltonian vector
field
\begin{equation}\label{}
  i\call_H = i\gamma ( n x^{n-1} p\partial_p - x^n\partial_x) \ ,
\end{equation}
does not define a self-adjoint operator for $n > 1$. In particular
for $n=2$ one obtains:
\begin{equation}\label{}
  e^{t \call_H} p = e^{2\gamma xt} p \ ,
\end{equation}
and
\begin{equation}\label{}
 e^{t \call_H} x = \gamma xt + (\gamma xt)^2 + (\gamma xt)^3 + \ldots \ ,
\end{equation}
which converges only for $|\gamma xt|< 1$. Quantization of
$H=x^np$ leads to $\widehat{H}  \propto \widehat{x}^n\widehat{p} +
\widehat{p}\widehat{x}^n$. It is well know (see e.g. \cite{Araki})
that this operator is not self-adjoint.

\section{Damped harmonic oscillator}

Consider a damped harmonic oscillator defined by the following set
of equations:
\begin{eqnarray}\label{}
  \dot{x} &=& - \gamma x + \omega p \ , \\
  \dot{p} &=& - \gamma p - \omega x\ .
\end{eqnarray}
Clearly this system is not Hamiltonian if $\gamma \neq 0$.
Applying the procedure of \cite{Pontr} one arrives at the
following Hamiltonian system on $ \mathbb{R}^4$ (for the physical
origin of this system cf. \cite{Kossak1}):
\begin{eqnarray}\label{}
\dot{x}^1 &=& \{ x^1,H\} = -\gamma x^1 + \omega x^2 \ , \\
\dot{x}^2 &=& \{ x^2,H\} = -\omega x^1 - \gamma x^2 \ , \\
\dot{p}_1 &=& \{ p_1,H\} = +\gamma p_1 + \omega p_2 \ , \\
\dot{p}_2 &=& \{ p_1,H\} = -\omega p_1 + \gamma p_2 \ ,
\end{eqnarray}
where we introduced $x^1=x$ and $x^2=p$. The Hamiltonian function
is given by:
\begin{equation}\label{H-osc-d}
  H(x,p) = \omega( p_1x^2 - p_2x^1) - \gamma
  (p_1x^1 + p_2x^2)\ .
\end{equation}
Introducing complex variables:
\begin{equation}\label{}
  z_1 = x^1 + ix^2\ \ \ \ \ {\rm and} \ \ \  \ \ z_2 = i(p_1 +ip_2)\ ,
\end{equation}
the Hamiltonian may be rewritten as follows:
\begin{equation}\label{}
  H(z_1,z_2,\overline{z}_1,\overline{z}_2) = \omega \, {\rm
  Re}(z_1\overline{z}_2) + \gamma\, {\rm Im}(z_1\overline{z}_2) \
  .
\end{equation}
The corresponding Koopman's operator
\begin{eqnarray}\label{}
  i\call_H &=& i\omega \left( x^2 \partial_{x^1} -
  x^1\partial_{x^2} + p_2\partial_{p_1} - p_1\partial_{p_2}
  \right) \nonumber\\ &+& i\gamma  \left( p_1\partial_{p_1} -
  x^1\partial_{x^1}  +  p_2\partial_{p_2} - x^2\partial_{x^2}\right)
  \ ,
\end{eqnarray}
takes  in the complex variables the following form:
\begin{equation}\label{Koopman-H-osc-d}
  i\call_H =  \left( \alpha\,  z_1 \partial_1 -
 \overline{\alpha}\,  \overline{z}_1\overline{\partial}_1
 \right) + \left( \overline{\alpha}\,z_2{\partial}_2  -
\alpha\,  \overline{z}_2\overline{\partial}_2 \right) \  ,
\end{equation}
where the complex parameter $\alpha$ is given by:
\begin{equation}\label{alpha}
  \alpha = \omega - i\gamma \ .
\end{equation}
The eigenvalue problem for $i\call_H$ is immediately solved by:
\begin{equation}\label{}
  i\call_H \, f_{nmkl} = \lambda_{nmkl}\, f_{nmkl} \ ,
\end{equation}
with
\begin{equation}\label{}
  f_{nmkl}(z_1,z_2) = z_1^n\overline{z}_1^mz_2^k\overline{z}_2^l \
  ,
\end{equation}
and
\begin{eqnarray}\label{nmkl}
  \lambda_{nmkl} &=& \alpha(n-l) - \overline{\alpha}(k-m) \nonumber \\
  &=&  \omega(n+k-m-l) - i\gamma(n+m-k-l) \ .
\end{eqnarray}
Evidently $f_{nmkl} \notin L^2( \mathbb{R}^4)$. However, it is
clear that the natural structure associated with the eigenvalue
problem is the following Gelfand triplet:
\begin{equation}\label{}
  {\cal S}( \mathbb{R}^4) \ \subset\ L^2( \mathbb{R}^4) \ \subset \ {\cal S}^*(
  \mathbb{R}^4) \ .
\end{equation}
Now, the spectrum given by (\ref{nmkl}) is discrete and complex.
Therefore, the classical resonant states $f_{nmkl}$ are
responsible for the damping of harmonic oscillations. The unitary
time evolution is given by:
\begin{equation}\label{}
  f_t(z_1,z_2,\overline{z}_1,\overline{z}_2) = e^{t\call_H}\,
  f(z_1,z_2,\overline{z}_1,\overline{z}_2) = f(e^{-i\alpha
  t}z_1,e^{-i\overline{\alpha}
  t}z_2,e^{i\overline{\alpha} t} \overline{z}_1,e^{i{\alpha} t} \overline{z}_2)
  \ .
\end{equation}
In particular
\begin{equation}\label{}
  z_1(t) = e^{-i\alpha t}z_1 = e^{-i\omega t} e^{-\gamma t} z_1\ .
\end{equation}
The quantized damped oscillator was analyzed in \cite{Kossak1}.
Due to (\ref{H-osc-d}) the quantum Hamiltonian reads:
\begin{equation}\label{H-osc-d-q}
  \widehat{H} = \omega( \widehat{p}_1\widehat{x}_2 - \widehat{p}_2\widehat{x}_1 ) -
  \frac{\gamma}{2} ( \widehat{p}_1\widehat{x}_1 + \widehat{x}_1\widehat{p}_1 +
  \widehat{p}_2\widehat{x}_2 + \widehat{x}_2\widehat{p}_2 ) \ .
\end{equation}
Introduce two sets of creaction and anihilation operators:
\begin{eqnarray}\label{}
  \widehat{a}_1 &=& \frac{\widehat{x}_1 + i\widehat{x}_2}{\sqrt{2\hbar}} \ , \
  \hspace{1cm}  \widehat{a}_1^* = \frac{\widehat{x}_1 -
  i\widehat{x}_2}{\sqrt{2\hbar}}\ , \\
 \widehat{a}_2 &=& \frac{i\widehat{p}_1 -\widehat{p}_2}{\sqrt{2\hbar}} \ , \
  \hspace{1cm}  \widehat{a}_2^* = \frac{-i\widehat{p}_1 -
  \widehat{p}_2}{\sqrt{2\hbar}}\ .
\end{eqnarray}
One finds that the following commutation relations hold:
\begin{equation}\label{CCR1}
  [\widehat{a}_1,\widehat{a}_2] = [\widehat{a}_1,\widehat{a}^*_1]
  = [\widehat{a}_2,\widehat{a}^*_2]= 0 \ ,
\end{equation}
\begin{equation}  \label{CCR2}
  [\widehat{a}_1,\widehat{a}_2^*] =  [\widehat{a}_2,\widehat{a}_1^*] = 1\
  ,
\end{equation}
and hence, the Hamiltonian (\ref{H-osc-d-q}) may be rewritten as
follows:
\begin{eqnarray}\label{H-a1-a2}
  \widehat{H} &=& \hbar\left(  \alpha\, \widehat{a}^*_2 \widehat{a}_1 +
  \overline{\alpha}\, \widehat{a}^*_1\widehat{a}_2 + \omega \right)
  \nonumber \\ &=&
 \hbar\alpha \left(  \widehat{a}^*_2 \widehat{a}_1 + \frac 12
 \right) +  \hbar\overline{\alpha}\left(  \widehat{a}^*_1 \widehat{a}_2 +
 \frac 12 \right) \ .
\end{eqnarray}
with $\alpha$ defined in (\ref{alpha}). Clearly, $\widehat{H}^* =
\widehat{H}$.
 Commutation relations (\ref{CCR1})--(\ref{CCR2}) may be represented
 in $C^\infty( \mathbb{R}^2)$ as follows \cite{Kossak1}:
\begin{eqnarray}\label{CCR1a}
\widehat{a}_1 &=& \frac{1}{\sqrt{2}} \left( x + \partial_y \right)
\ , \hspace{1cm} \widehat{a}^*_1 = \frac{1}{\sqrt{2}} \left( x -
\partial_y \right) \ , \\   \label{CCR2a}
\widehat{a}_2 &=& \frac{1}{\sqrt{2}} \left( y + \partial_x \right)
\ , \hspace{1cm} \widehat{a}^*_1 = \frac{1}{\sqrt{2}} \left( y -
\partial_x \right) \ .
\end{eqnarray}
Inserting (\ref{CCR1a})--(\ref{CCR2a}) into (\ref{H-a1-a2}) one
obtains the following representation for $\widehat{H}$:
\begin{equation}\label{}
  \widehat{H} = \hbar \omega \left( xy - \partial^2_{xy} \right) -
  i\hbar \gamma \left( y\partial_y - x\partial_x \right)\ .
\end{equation}
The spectrum oh $\widehat{H}$ is easy to find. One defines a
``ground state'' $\varphi_{00}$ to be the state satisfying
\begin{equation}\label{phi-00}
  \widehat{a}_1 \varphi_{00} = \widehat{a}_2\varphi_{00} =0\ ,
\end{equation}
and
\begin{equation}\label{}
\varphi_{nk} := (\widehat{a}^*_1)^n(\widehat{a}^*_2)^k
\varphi_{00}\ .
\end{equation}
Now, it is easy to show that
\begin{equation}\label{}
  \widehat{H} \varphi_{nk} = \mu_{nk} \varphi_{nk} \ ,
\end{equation}
with
\begin{equation}\label{mu-nk}
 \mu_{nk} = \hbar\alpha\left(n+ \frac 12 \right) +
 \hbar\overline{\alpha}\left( k + \frac 12 \right)
 = \hbar \omega (n+k+1) - i\hbar\gamma(n-k)\ .
\end{equation}
Solving (\ref{phi-00}) one obtains (up to a non-important
constant):
\begin{equation}\label{}
  \varphi_{00}(x,y) = e^{-xy}\ ,
\end{equation}
and hence the eigenstates $\varphi_{nk}$ do not belong to $L^2(
\mathbb{R}^2)$.

\par
\noindent {\bf Remark 3.} Let us observe that commutation
relations (\ref{CCR1})--(\ref{CCR2}) may be represented in the
space of holomorphic functions of two complex variables equipped
with the following scalar product:
\begin{equation}\label{}
  \la \psi|\phi\r_ = \int \overline{f}(z_1,z_2)g(z_1,z_2)
  e^{-z_1\overline{z}_2 - \overline{z}_1z_2}\, dz_1dz_2d\overline{z}_1d\overline{z}_2\ .
\end{equation}
One has:
\begin{eqnarray}\label{}
  \widehat{a}_1 &=& \partial_1\ , \hspace{1cm} \widehat{a}_1^* = z_2\ , \\
\widehat{a}_2 &=& \partial_2\ , \hspace{1cm} \widehat{a}_2^* =
z_1\ .
\end{eqnarray}
Note, that contrary to (\ref{a-a*}) the above representation is
not consistent with the corresponding Bargmann scalar product,
that is, $\widehat{a}^*_k$ is not the adjoint of $\widehat{a}_k$
with respect to (\ref{B-product}). Now, using (\ref{H-a1-a2}) one
obtains:
\begin{equation}\label{hat-H-zdz}
  \widehat{H} = \hbar \left( \alpha z_1 \partial_1 + \overline{\alpha}
  z_2\partial_2   + \omega  \right) \ ,
\end{equation}
which is the quantum analog of the Koopman's operator
(\ref{Koopman-H-osc-d}). The eigenvalue problem for
(\ref{hat-H-zdz}) is solved by:
\begin{equation}\label{}
  \widehat{H} \psi_{nk} = \mu_{nk}\, \psi_{nk}\ ,
\end{equation}
with
\begin{equation}\label{}
  \psi_{nk}(z_1,z_2) = z_1^nz_2^k\ ,
\end{equation}
and $\mu_{nk}$ are given by(\ref{mu-nk}). Clearly, holomorphic
eigenvectors $f_{n0k0}= \psi_{nk}$ and anti-holomorphic $f_{0m0l}
= \overline{\psi}_{ml}$.

\section{Concluding remarks}

Let us observe that  introducing polar coordinates $(r,\varphi)$
on $ \mathbb{R}^2$, one finds that $i\call_{H_{\rm osc}}$ defines
a generator of $SO(2)$ rotation:
\begin{equation}\label{d-phi}
  i\call_{H_{\rm osc}} = i\omega \partial_\varphi \ .
\end{equation}
Therefore, the eigenvalue problem for $i\call_{H_{\rm ocs}}$ may
be translated to the corresponding problem for $i\partial_\varphi$
with an obvious solution given by:
\begin{equation}\label{}
  i\omega\partial_\varphi\, e^{-im\varphi } = m\omega \,
  e^{-im\varphi }\ ,
\end{equation}
with $m\in \mathbb{Z}$. Hence, one recovers (\ref{Koopman-Osc}).
Note, however, that we have reduced the problem from $L^2(
\mathbb{R}^2)$ to $L^2(S^1)$, i.e. we consider a representation of
$i\call_{H_{\rm osc}}$ on a compact space $S^1$, which is a
homogeneous space for $SO(2)$.

Now, let us turn to the  damped system described by (\ref{H-XP}).
Introducing hiperbolic coordinates:
\begin{equation}\label{s-chi}
  P = s\cosh\chi \ \ \ \ {\rm and} \ \ \ \ X=s\sinh\chi \ ,
\end{equation}
with $s\in (-\infty,\infty)$ and $\chi \in
(-\infty,\infty)$\footnote{Clearly, this parameterization covers
only the interior of the ``light cone'' $|P|>|X|$. To cover also
the remaining region one chooses $P=s'\sinh\chi'$ and
$X=s'\cosh\chi'$.} one easily finds that
\begin{equation}\label{d-chi}
  i\call_H = i\gamma \partial_\chi\ ,
\end{equation}
which defines a generator of $SO(1,1)$.  Note however, that
contrary to the generator of the compact group $SO(2)$, the
corresponding generator of a non-compact $SO(1,1)$ has a
continuous spectrum:
\begin{equation}\label{}
i\gamma \partial_\chi\, e^{\beta\chi} = i\gamma\beta\,
e^{\beta\chi} \ ,
\end{equation}
with $\beta \in \mathbb{R}$. But this is not what we have found in
(\ref{Spec-c}). This example shows  that changing representation
we may change the spectrum of the corresponding operator.
Therefore, one has to be careful choosing the appropriate
representation which has to be dictated by the physical problem in
question. Actually, our problem is defined on $ \mathbb{R}^2$, and
hence, the corresponding Gelfand triplet ${\cal S}( \mathbb{R}^2)
\subset L^2( \mathbb{R}^2) \subset {\cal S}^*(\mathbb{R}^2)$ leads
to the discrete spectrum.

Finally, let us note  that the spectrum of a self-adjoint operator
$A$ on a Hilbert space $\cal H$ does depend upon the space of test
function $D$ in the definition of the Gelfand triplet
(\ref{triplet}). Consider for example a momentum operator
$\widehat{p} =-i\partial_x$. Clearly $\widehat{p}$ has purely real
eigenvalues $p\in \mathbb{R}$ corresponding to plane waves $\psi_p
= e^{ipx}$:
\begin{equation}\label{}
  \widehat{p}\, \psi_p =  p \psi_p\ ,
\end{equation}
but it has also complex eigenvalues $z$:
\begin{equation}\label{}
  \widehat{p}\, \psi_z = z\, \psi_z\ .
\end{equation}
with $\psi_z = e^{-iz x}$. Obviously neither $\psi_k$ nor $\psi_z$
belong to $L^2( \mathbb{R})$. Every student accepts ``$p$'' but
``$ z$'' looks quite strange. However, mathematically they may be
treated on the equal footing.  It is evident that $\psi_p,\psi_z
\in D^*( \mathbb{R})$, where $D( \mathbb{R})$ is the space of
smooth function with compact supports equipped with the convex
Schwartz topology (cf. e.g. \cite{Yosida}).  Now, consider two
subspaces $D_+( \mathbb{R})$ and $D_-( \mathbb{R})$ in $C^\infty(
\mathbb{R})$:
\begin{equation}\label{}
D_+( \mathbb{R})   := \Big\{ \ \phi \in C^\infty( \mathbb{R})\ |\
{\rm
  supp}\,\phi \subset (-\infty,c)  \ \Big\}  \ ,
\end{equation}
\begin{equation}\label{}
D_-( \mathbb{R}) := \Big\{ \ \phi \in C^\infty( \mathbb{R})\ |\
{\rm
  supp}\,\phi \subset (c,\infty)  \ \Big\}  \ ,
\end{equation}
for some $c \in \mathbb{R}$. Note, that
\[   \psi_z \ \in \  \left\{  \begin{array}{ll}
D^*_+( \mathbb{R}) \ , &\ \ \ \ {\rm if} \ \ {\rm Im}\,z
> 0 \\
 D^*_-( \mathbb{R}) \ , &\ \ \ \ {\rm if} \ \ {\rm Im}\,z < 0\ \end{array} \right.  , \]
but $\psi_p$, which corresponds to ${\rm Im}\,z=0$, does not
belong to $D^*_\pm( \mathbb{R})$ for any $p$.

\vspace{.5cm}

The author thanks Professor Andrzej Kossakowski for very
interesting and stimulating discussions.  This work was partially
supported by the Polish State Committee for Scientific Research
(KBN) Grant no 2P03B01619.

\end{document}